\documentclass[reprint,aps,pra,amssymb,amsmath,superscriptaddress,showpacs,floatfix]{revtex4-1}
\usepackage{times}
\usepackage{graphicx}
\usepackage{tabularx}
\usepackage{subfigure}
\usepackage{dcolumn}
\usepackage{bm}
\usepackage{multirow}

\newcommand{\wavenum}{cm$^{-1}$}

\newcommand{\X}{X$^1\Sigma^+$}
\newcommand{\AAA}{A$^1\Sigma^+$}
\newcommand{\C}{C$^1\Sigma^+$}
\newcommand{\A}{a$^3\Sigma^+$}
\newcommand{\CC}{c$^3\Sigma^+$}
\newcommand{\BB}{b$^3\Pi$}
\newcommand{\B}{B$^1\Pi$}
\newcommand{\D}{D$^1\Pi$}
\newcommand{\Three}{$(3)^3\Pi$}
\newcommand{\Four}{$(4)^3\Sigma^+$}
\newcommand{\Fo}{$(4)^1\Sigma^+$}
\newcommand{\Fi}{$(5)^1\Sigma^+$}

\begin{document}

\preprint{APS/123-QED}

\title{Formation of deeply bound ultracold LiRb molecules via photoassociation near the Li 2S$_{1/2}$ + Rb 5P$_{3/2}$ asymptote}

\author{John Lorenz}
\email{lorenz.john@gmail.com}
\affiliation{Department of Physics, Purdue University, West Lafayette, Indiana 47907, USA}
\author{Adeel Altaf}
\affiliation{Department of Physics, Purdue University, West Lafayette, Indiana 47907, USA}
\author{Sourav Dutta}
\affiliation{Department of Physics, Purdue University, West Lafayette, Indiana 47907, USA}
\affiliation{Raman Research Institute, C. V. Raman Avenue, Sadashivanagar, Bangalore - 560080, India}
\author{Yong P. Chen}
\affiliation{Department of Physics, Purdue University, West Lafayette, Indiana 47907, USA}
\affiliation{School of Electrical and Computer Engineering, Purdue University, West Lafayette, Indiana 47907, USA}
\author{D. S. Elliott}
\email{elliottd@purdue.edu}
\affiliation{Department of Physics, Purdue University, West Lafayette, Indiana 47907, USA}
\affiliation{School of Electrical and Computer Engineering, Purdue University, West Lafayette, Indiana 47907, USA}

\begin{abstract}
We present spectra of ultracold $^7$Li$^{85}$Rb molecules in their electronic ground state formed by spontaneous decay of weakly bound photoassociated molecules. Beginning with atoms in a dual species magneto-optical trap (MOT), weakly bound molecules are formed in the 4(1) electronic state, which corresponds to the \B\ state at short range.  These molecules spontaneously decay to the electronic ground state and we use resonantly enhanced multiphoton ionization (REMPI) to determine the vibrational population distribution in the electronic ground states after spontaneous emission.  Many of the observed lines from the spectra are consistent with transitions from the \X\ ground electronic state to either the \B\ or \D\ electronic states that have been previously observed, with levels possibly as low as \X\ $(v'' = 2)$ being populated.  We do not observe decay to weakly bound vibrational levels of the \X\ or \A\ electronic states in the spectra.  We also deduce a lower bound of 3900 \wavenum\ for the dissociation energy of the LiRb$^+$ molecular ion.
\end{abstract}


\maketitle

There has been much attention given to cold polar molecules~\cite{carr09, ulmanis12} due to their potential as a medium for precision measurements~\cite{acme14, hudson11, leanhardt11}, quantum computation~\cite{demille02}, quantum simulation~\cite{barnett06}, investigations of time variation in fundamental constants~\cite{flambaum07}, and ultracold quantum chemistry~\cite{ospelkaus10, zuchowski10}. Experiments that utilize the internal structure of a molecule can be conducted using molecular ions or neutral molecules. However, experiments that seek to take advantage of the dipole-dipole interactions require an ultracold sample of neutral polar molecules~\cite{demille02}. Their dipole moment, and thus interaction strength, is strongest in the rovibronic ground state. The intrinsic lifetime of a molecule in its ground state is infinite making it an ideal quantum state for studies related to quantum information, computation or simulation (in practice, however, the lifetime is limited by the quality of vacuum, depth of the molecular trap etc.) So far, the most straightforward way to produce ultracold molecules is through photoassociation (PA) or magnetoassociation (MA) of ultracold atoms in a trap. PA involves inducing a scattering-bound transition via an optical field~\cite{lett93}, while MA involves tuning an external magnetic field to a Feshbach resonance~\cite{kohler06}. Because alkali metals have a single valence electron and a strong cycling transition, they are commonly used in ultracold atomic traps and the most commonly studied ultracold molecules are bi-alkalis.  Among the bi-alkalis, LiRb is attractive due to its high PA rate (highest among the bi-alkalis~\cite{dutta14,dutta13}), and a large permanent electric dipole moment in the rovibronic ground state (third highest among bi-alkalis ~\cite{aymar05}).
Ultracold Li-Rb mixtures are also a possible candidate for study of a gas-crystal quantum transition~\cite{gacesa13,petrov07}.  However, until recently, LiRb had been one of the least experimentally studied bi-alkali molecules. Recent heat pipe spectra~\cite{dutta11,ivanova11,ivanova13}, measurements of large Feshbach resonances~\cite{deh08,marzok09,deh10}, collision studies~\cite{duttacollision14}, PA spectra~\cite{dutta14, dutta13}, and ionization spectra of the \A\ triplet ground state potential~\cite{altaf}, have given a much clearer picture of the LiRb molecular structure and how Li and Rb interact.  Prior knowledge of LiRb was based solely on \textit{ab initio} calculations~\cite{korek00,korek09}.  In this article, we present resonantly enhanced multiphoton ionization (REMPI) spectra of ultracold LiRb molecules in their electronic ground state. Such molecules are formed by spontaneous decay of weakly bound LiRb* molecules produced by PA to vibrational levels just below the Li 2S$_{1/2}$ + Rb 5P$_{3/2}$ asymptote.  We observe that excited LiRb$^*$ molecules  formed via PA decay to deeply bound levels of the \X\ potential via spontaneous emission (SE), possibly populating as low as the $v'' = 2$ level.  (Throughout this paper, we use an asterisk to denote an excited electronic state, $v''$ to denote the vibrational levels the ground electronic state, $v$ (without a prime) to denote the vibrational levels of any of the PA resonances, and $v'$ to denote the vibrational level of other excited electronic states.)  We see a number of spectral features/progressions consistent with population in these levels ionizing through either the \B\ or \D\ potentials. Contrary to expectation, we see no evidence that LiRb$^*$ molecules created near the dissociation limit decay to weakly-bound vibrational levels of the \X\ or \A\ states.

The details of our experimental apparatus can be found in previous reports from our group~\cite{duttacollision14,dutta14,altaf}. We have a dual species $^7$Li/$^{85}$Rb MOT typically trapping $5 \times 10^7$ Li atoms and $1 \times 10^8$ Rb atoms at densities of $5 \times 10^9$ cm$^{-3}$ and $4 \times 10^9$ cm$^{-3}$, respectively.  We use a dark SPOT~\cite{ketterle93} for Rb to minimize losses due to collisions between Li and Rb$^*$~\cite{duttacollision14}.  We photoassociate LiRb using a Coherent 899-21 Ti:Sapphire laser at powers between 350--500 mW and a beam diameter of 0.85 mm.  To detect ground state LiRb molecules after SE, we rely on REMPI. The ionization laser is a Quanta-Ray PDL-2 Pulsed Dye Laser (PDL) pumped by a Quanta Ray PRO-Series Pulsed Nd:YAG laser. We use Rhodamine 590 and Rhodamine 610 dyes to generate light in the wavelength range of 556--610 nm, corresponding to frequencies in the range of 16400--18000 \wavenum . The pulse width is about 7 ns and we generally use pulses of 2-3 mJ in energy to induce two-photon ionization.  The repetition rate of the laser is 10 Hz. The frequency of the dye laser is measured with a wavelength-meter and is determined with a typical accuracy of 0.2 \wavenum . Because this laser pulse can ionize Li, Rb, LiRb, and Rb$_2$, we rely on a time-of-flight (TOF) mass spectrometer in our vacuum chamber to separate the different ionic species that are incident on our micro-channel plate (MCP) detectors. We use a discriminator to monitor the amplified signal from the MCP anode and count the number of ions arriving within the LiRb TOF window that is synchronized to each laser pulse. The MOT coils and TOF field plates are left on for the duration of the experiment. 

\begin{figure}
	\includegraphics[width=.48\textwidth, clip=true]{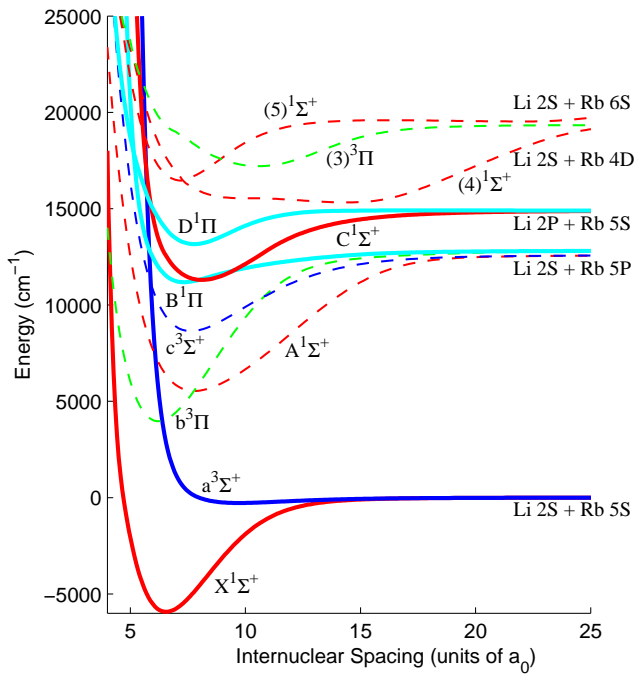}
	\caption{\label{PECs} (Color online) LiRb potential energy curves potentially involved in the PA or REMPI processes. Molecules are formed via PA to the $4(1)$ state (\B) close to the Rb 5P$_{3/2}$ asymptote. Molecules accumulating in the ground state (\A\ or \X) have several REMPI pathways depending on the binding energy of the ground state vibrational level. For REMPI frequencies 16,400--18,000 \wavenum , \X\ molecules populating $v'' < 27$ ionize through the \B , \C , or \D\ potentials. Higher $v''$ states would ionize through the \Fo\ or \Fi\ potentials. Molecules decaying to the \A\ state ionize primarily through the \Three\ potential.}
\end{figure}

In Fig.~\ref{PECs}, we show the potential energy curves (PECs) of the states involved in the PA and REMPI processes.  The solid PECs are based on Fourier-transform spectroscopy of LiRb produced in a heat pipe by Ivanova \textit{et al.}\ ~\cite{ivanova11,ivanova13}. The dashed curves are the results of  \textit{ab initio} calculations from Korek \textit{et al.}~\cite{korek09}. In our previous experiments, we detected a number of PA resonances in the $2(0^+)$, $2(0^-)$ and $2(1)$ potentials below the Rb 5P$_{1/2}$ atomic resonance~\cite{dutta13,altaf} and in the $3(0^+)$, $4(1)$, and $1(2)$ potentials below the 5P$_{3/2}$ asymptote~\cite{dutta14,altaf}. 
We use Hund's case (c) notation to denote the long range potentials. Their correspondence to short range Hund's case (a) or (b) states can be found in Table~\ref{hunds}. Previous REMPI scans taken after PA to the $2(0^-)$ state allowed us to observe the \A , \Three , and \Four\ potentials~\cite{altaf}. 

\begin{table}
  \caption{\label{hunds} Correspondence between short range electronic states--Hund's case (a) or (b) and long range--Hund's case (c) and asymptotic atomic states in LiRb. PECs are considered diabatically.}
  \begin{ruledtabular}
  \begin{tabular}{ c  c  c }
       Hund's case (a)/(b) &  Hund's case (c) & Asymptote \\ \hline
       \X & 1(0$^+)$ & \multirow{4}{*}{Li 2S$_{1/2}$ + Rb 5S$_{1/2}$} \\ \\
       \multirow{2}{*}{\A} & 1(0$^-)$ \\ & 1(1) \\ \hline
       \AAA & 2(0$^+$) & \multirow{4}{*}{Li 2S$_{1/2}$ + Rb 5P$_{1/2}$} \\ \\
       \multirow{2}{*}{\CC} & 2(0$^-$) \\ & 2(1) \\ \hline
       \multirow{4}{*}{\BB} & 3(0$^+$) & \multirow{6}{*}{Li 2S$_{1/2}$ + Rb 5P$_{3/2}$} \\
       & 3(0$^-$) \\ & 3(1) \\ & 1(2) \\ \\ 
       \B & 4(1) \\
  \end{tabular}
  \end{ruledtabular}
    
\end{table}

\begin{figure*}
	\includegraphics[width=\textwidth, clip=true]{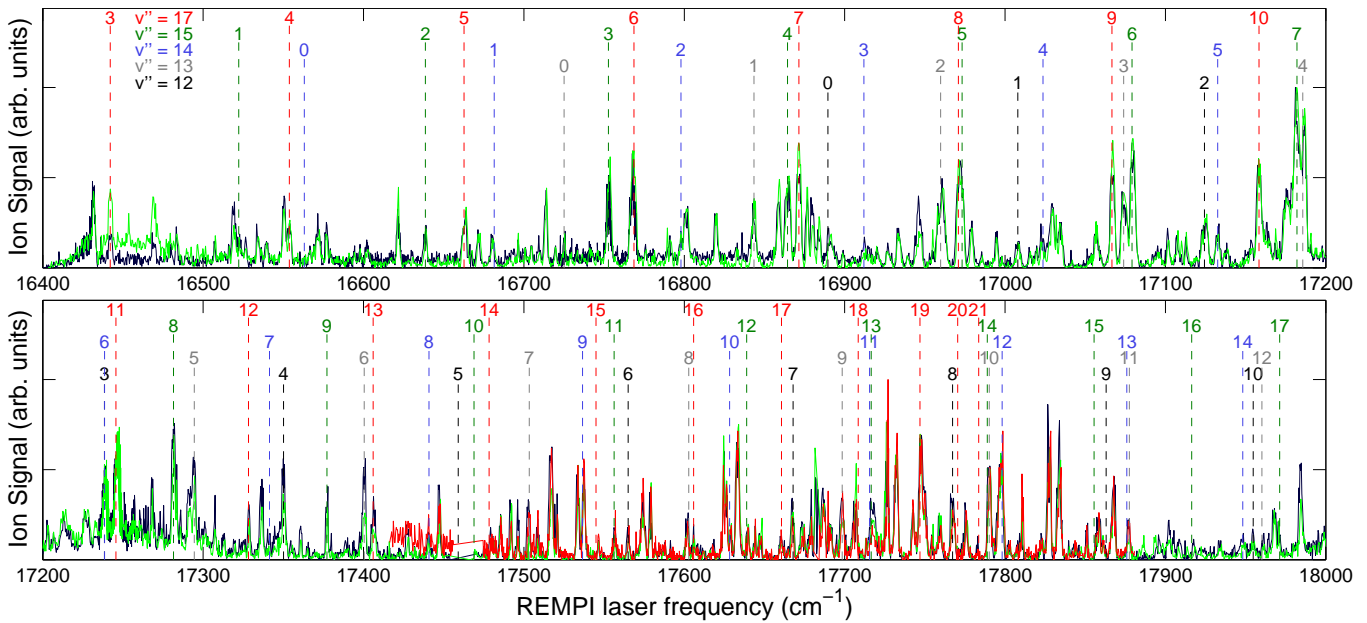}
	\caption{\label{spectra} (Color online) REMPI spectra with transitions from \X\ levels ($v''$) to \D\ levels (numbered above dashed line) overlaid. The blue (dark gray), green (light gray), and red (gray, only present in bottom panel) data represent REMPI spectra taken while the PA laser is resonant with the $(v = -3, J = 2)$, $(v=-4, J=1)$, and $(v=-5, J=1)$ states of the $4(1)$ potential, respectively. Data has been normalized so that all spectra are visible on the same plot. For clarity, only five of the established progressions are shown.}
\end{figure*}

In this work, we present REMPI spectra recorded while the PA laser is resonant with the $v = -3,-4,$ or $-5$ levels of the $4(1)$ potential, which we observed through previous trap-loss and ionization spectroscopy~\cite{dutta14,altaf}. (Vibrational states of PA resonances are labeled using negative values relative to the dissociation limit, with $v = -1$ being the least bound vibrational state. Labeling vibrational levels with respect to the most deeply bound vibrational level requires additional spectroscopy to connect observations of long range states to previously measured short range states in LiRb.)  To begin labeling spectral lines, we considered possible contributions from many different molecular levels.  For example,  4(1) PA states can decay to the $1(1)$, $1(0^-)$, or $1(0^+)$ ground states under Hund's case (c) selection rules~\cite{herzberg}. Since these correspond to ground triplet or singlet states at short range (see Table~\ref{hunds}) we anticipated population in both ground states following PA. (In contrast, the parity selection rule allows the $2(0^-)$ state to decay only to $1(1)$ or $1(0^-)$, both corresponding to the triplet state.) 
Thus many ionization pathways are possible depending on how molecules decay after PA from the 4(1) state.  Moreover, molecules that decay to \X\ levels can ionize via a two-photon process through a number of intermediate-state potentials, depending on how deeply bound they are.  The \X\ potential has a depth of 5928(4) \wavenum ~\cite{ivanova11}.  Very little is known regarding the states of the LiRb$^+$ ion.  Based on the calculations of von Szentp\'aly \textit{et al.}~\cite{vonszentpaly82} and Azizi~\cite{azizi}, which were supported by REMPI observations from LiCs experiments~\cite{deiglmayrfaraday09}, we anticipate that $D_e$ and $\omega_e$ of the X$^2\Sigma^+$ potential of the LiRb$^+$ ion are close to 4200 \wavenum\ and 140 \wavenum , respectively. 
Based on the known or estimated PECs for LiRb, we anticipated that PDL frequencies in the range of 16,400--18,000 \wavenum\ should ionize molecules in the $v'' = 0$--10 levels of \X\ via two-photon REMPI through \B\ states, and molecules in the $v'' = 7$--26 can be ionized through \D\ states.  For $v'' > 26$, molecules would have to ionize through higher potentials such as \Fo\ or \Fi . Molecules occupying levels $v'' \leq 26$ could also ionize through the \C\ state.
The bottom of the \C\ potential is known indirectly  
through its perturbations of 
states belonging to the \B\ potential~\cite{ivanova13}, but only a few levels in the known range could participate in REMPI. Thus, we focus on transitions involving vibrational levels belonging to the \B\ and \D\ potentials. 

Fig.~\ref{spectra} shows REMPI spectra that we have generated via PA to $(v = -3, J = 2)$, $(v=-4, J=1)$, and $(v=-5, J=1)$ states of the $4(1)$ potential. We have normalized these spectra so that they are visible on the same plots, regardless of PA rate.  
The $(v = -3, J = 2)$ PA transition is the strongest, with these data being integrated for 50 laser pulses and yielding about 2.5 LiRb$^+$ ions per pulse on the strongest line.  The $(v=-4, J=1)$ and $(v=-5, J=1)$ are weaker PA transitions that required integration over 200 and 250 laser pulses, respectively.  The maximum count rate we observed for these weaker PA resonances was 1.0 and 0.4 ions per pulse, respectively. The choice of $J$ for a given $v$ was simply based on which PA resonance resulted in a stronger REMPI signal. While we scanned the PDL between 16,400--18,000 \wavenum\ for the $v = -3$ and $-4$ resonances, we scanned over a smaller region of 17,425--17,875 \wavenum\ for the $v=-5$ resonance.  We limited the scan range for the $v=-5$ PA resonance because of the large integration time required and the fact that the normalized spectra for $v = -5$ is nearly identical to the $v = -3$ and $-4$ spectra, indicating each of the three PA resonances has very similar decay pathways. Note that there are a couple of gaps in the scans which correspond to atomic Rb resonances.  The 5P$_{3/2} \rightarrow$ 7D$_{5/2, 3/2}$ transitions occurs at 17463.57 \wavenum\ and 17465.08 \wavenum\ and the 5P$_{3/2} \rightarrow$ 9S$_{1/2}$ transition occurs at 17682.49 \wavenum\ \cite{NIST_ASD}.  These electric dipole-allowed transitions result in strong two-photon ionization of the excited Rb atoms.  Approaching these energies on a REMPI scan results in a much stronger Rb$^+$ signal on the mass spectrometer that gives false positives in our LiRb$^+$ detection window.  Therefore, we omit these regions from our spectra.

\begin{figure*}
	\includegraphics[width=\textwidth, clip=true]{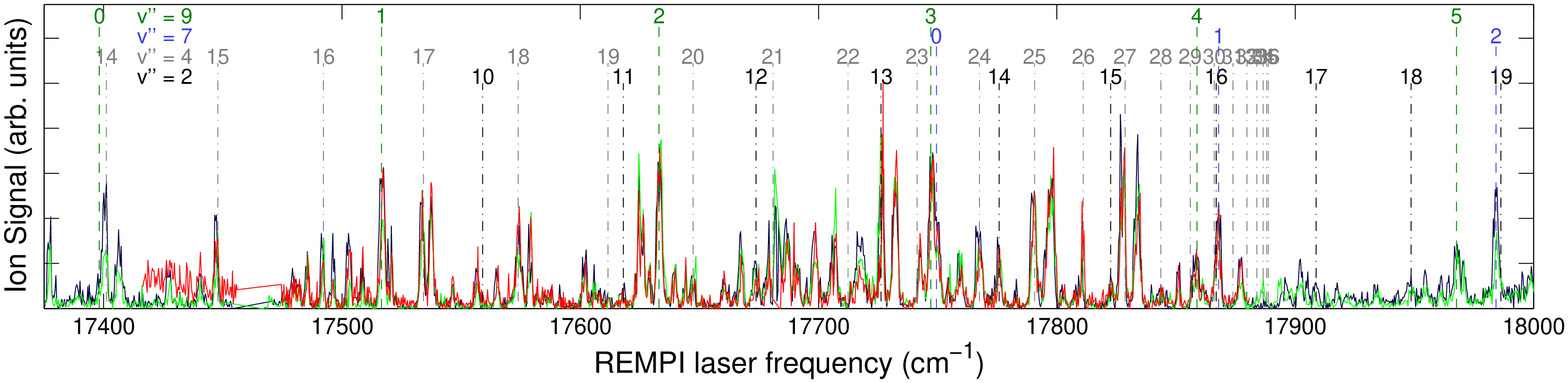}
	\caption{\label{B} (Color online) REMPI spectra between 17,400--18,000 \wavenum . Apparent progressions are from the deeply bound $v'' = 2,4,7,9$ levels of the \X state. The $v'' = 2,4$ progressions ionize through \B\ levels, indicated by dot-dashed lines. The $v'' = 7,9$ ionize through \D\ levels, indicated by dashed lines. Progressions originating from $v'' = 3,5$ through \B\ levels are omitted for clarity. The $v'' = 2$ progression represents the most deeply bound population observed.}
\end{figure*}

The spectra contain information about the distribution of population in the ground state after SE and the excited states involved in REMPI, PECs which can be found in Fig.~\ref{PECs}. The PECs calculated from heat pipe data (solid lines) should be quite accurate. Specifically, the $v' = 0$--10, 13, and 15 vibrational levels of the \D\ have been directly observed~\cite{ivanova13}.  Unobserved levels between $v' = 11$--16 should be well known as well, because the PEC used to calculate their energy is known to fit nearby vibrational levels well. We used the public distribution of LEVEL 8.0~\cite{level} to calculate the positions of the unobserved levels up to $v' = 21$ for the \D\ potential based on the PECs reported~\cite{ivanova13}. Comparing the spectra in Fig.~\ref{spectra} to known \X\ $\rightarrow$ \D\ transitions, a number of progressions that originate from population in the $v'' = 7,9$--$15,17$--$20$, and 22 levels of the \X\ potential can be seen. Five of the strongest progressions are shown in the figure. The strongest line observed corresponds to the $v'' = 15 \rightarrow v' = 7$ transition. Transitions for $v'' = 15$ to \D\ levels can be seen from $v' = 1$ to $v' = 13$ in the spectra with line intensities steadily decreasing on either side of $v' = 7$ transition. We establish progressions by observing sequences of lines consistent with transitions to several consecutive excited vibrational levels appearing in the spectra. 
Due to the number of possible transitions in this range and the spectral linewidth of the peaks, an observed line may correspond to multiple unresolved molecular transitions.  The presence of adjacent transitions can often resolve this issue. We have rejected some apparent assignments because of an inconsistent progression. For example a few lines could be assigned to $v'' = 16$, but there is no clear progression of consecutive vibrational levels.

In our search for evidence of population in  low vibrational levels of the \X\ state, we focus on REMPI transitions through intermediate \B\ levels. In Fig.~\ref{B} we have indicated transitions from the $v'' = 2$ and $4$ levels of the \X\ state to \B\ levels.  (The data in this figure is repeated from~\ref{spectra}.)  We have also identified possible progressions originating from the $v'' = 3$ and $5$, but these are omitted from the figure for clarity. 
The $v'' = 2$ vibrational level is the most deeply bound state that we observe in our spectra.  In Fig.~\ref{B}, the $v'' = 2 \rightarrow v' = 11$--15 transitions can be seen, although $v' = 11$ and 15 are weak. Peaks for $v' = 16$--18 may also be present, but are not sufficiently clear to allow us to identify these with certainty.
\B\ vibrational levels have been observed for $v' < 23$~\cite{dutta11,ivanova13} and a potential for the \B\ state, extending up to its dissociation limit, has been derived based on them. The derived PEC is thus not expected to be very accurate for $v' > 23$. Nevertheless, we used LEVEL to calculate the vibrational levels for $v \geq 23$ and observe lines that could correspond to $v'' = 4 \rightarrow v' = 23$--27, as seen in Fig.~\ref{B}. These assignments are only tentative, and efforts in our laboratory to observe more deeply bound PA resonances that are part of this same progression should help clarify these. Observation of the $v = -3,-4$, and $-5$ levels of the $4(1)$ potential via REMPI would have allowed us to greatly reduce the uncertainty in the \B\ PEC and helped to identify high $v'$ states as well, but these lines were not present in our spectra.  
A possible explanation is that there is insufficient FC overlap in the second step of the REMPI process for a two-photon transition to occur.

Prior to starting these measurements, we expected that there should be some decay from the $4(1)$ state to the $1(0^-)$ or $1(1)$ states (\A\ at short range). However, we were unable to observe any lines originating from the \A\ ground state \cite{altaf}.  
Similarly, we expected to observe population of weakly-bound vibrational levels of the \X\ ground state following PA, due to strong FC overlap with the PA level. 
As discussed earlier, for the wavelength range of our pulsed laser, vibrational levels of the \Fo\ potential would provide the intermediate resonance for REMPI from these states,  
manifested in the spectra as a pattern of lines 
repeating at the vibrational energy spacing of the \Fo\ potential --- about 40--45 \wavenum\ over a large range based on the calculated PECs~\cite{korek09,level}. We could not find such a pattern, indicating that either these weakly-bound ground state levels are not populated, or the REMPI cross section is very small. Formation of deeply bound molecules via photoassociation near the dissociation limit has been presumed based on theoretical models in LiK and RbCs experiments~\cite{ridinger11,kerman04} and directly observed in NaCs~\cite{zabawa10}. However, the lack of evidence of a weakly-bound ground state population is not well understood and perhaps requires a better understanding of the LiRb potentials to label unassigned lines in the spectra.

The lines that we observe also give the first experimental information about the X$^2\Sigma^+$ ground state potential of the LiRb$^+$ ion. By comparing the energy of the Li + Rb$^+$ asymptote (33,690.8 \wavenum ~\cite{NIST_ASD}) to the final energy of the LiRb$^+$ ion produced by the REMPI process (the two-photon energy less the binding energy of the initial state), a minimum value for the binding energy of the final state of the ion can be established.  This energy must lie at least $\frac{1}{2}\omega_e$ above the bottom of the potential. The \X\ $v'' = 15 \rightarrow$ \D\ $v' = 1$ line establishes a minimum $D_e$ of 3900 \wavenum . Alternatively, the weak \X\ $v'' = 2 \rightarrow$ \B\ $v' = 11$ line yields a minimum depth of 3970 \wavenum , but the assignment of this line is less certain, and we quote the more conservative estimate of $D_e \geq 3900$ \wavenum . Similar to REMPI spectra of LiCs~\cite{deiglmayrfaraday09}, this favors the previously mentioned calculations from Refs.~\cite{vonszentpaly82,azizi} over calculations from Patil and Tang~\cite{patil00}. The calculated potential depth reported by Ghanmi \textit{et al.}~\cite{ghanmi12} is not excluded by the conservative estimate, but would be by any stronger constraint.

\begin{figure}
	\includegraphics[width=.48\textwidth, clip=true]{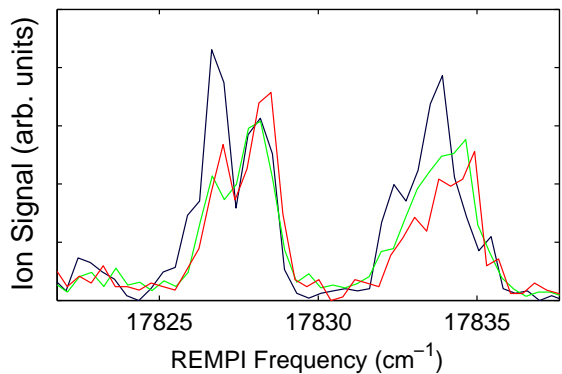}
	\caption{\label{zoom} (Color online) Example of differences between REMPI scans after PA to $J=2$ state (blue/dark gray) and $J=1$ (green/light gray and red/gray). The left peak exhibits a doublet where the intensities differ based on the rotational state of the PA transition involved. On the right the $J=1$ REMPI scan exhibits a line shape that is slightly shifted in peak intensity.}
\end{figure}

We conclude by noting that many of the ambiguous or uncertain assignments in the spectra could perhaps be settled through improved frequency resolution of the spectra, such that the rotational levels of the molecule could be resolved.   This could be implemented in our set-up, for example, by depleting the ground state population with a tunable narrow-band laser, while measuring the REMPI signal on a fixed line~\cite{deiglmayr08,zabawa11}. The rotational constant, $B_v$, is unique to each vibrational level and could be used to identify the vibrational level of the ground state molecule being ionized. While rotational levels are unresolved in the REMPI spectra, there do appear to be rotational effects that depend on $J$ of the excited molecule.  In Fig.~\ref{zoom}, we show an expanded view of a narrow range of the spectrum, in which the lineshapes differ depending on whether molecules are initially formed in a $J=2$ or $J=1$ state.  The maxima of the peaks shift depending on the rotational level used for PA. The shift can be subtle for some peaks, but quite pronounced as some peaks start to split into doublets of different intensities. However the effect manifests, it is reproducible when comparing PA resonances with different values for $J$. Because the initial rotational state is not expected to affect the vibrational level population after SE, it is likely that this effect is due to a difference in rotational population after SE. On the peak to the left (Fig.~\ref{zoom}), this shift starts to resolve into a doublet. The spacing between the peaks of the doublet is $>$ 1 \wavenum . Such a large spacing between features that arise due to rotational population is another indication of deeply bound ground state molecules.

The LiRb spectra presented in this work demonstrate that deeply bound LiRb molecules can be formed through PA to vibrational levels near the dissociation limit in the excited state, where the PA rate is quite large. The ground state molecular formation rate after PA to the $v = -3$ level of the $4(1)$ state is comparable to what was seen in PA to the $2(0^-)$ state --- in the range of $10^5$--$10^7$ s$^{-1}$. The formation rate for PA to the $v = -5$ state is roughly an order of magnitude lower.  Below the $v = -5$ state, the molecular formation rate appears to drop to barely detectable levels. Our ultimate goal is to find an efficient path to the rovibronic ground state. Currently, the spectra indicate that we are able to populate as low as $v'' = 2$ simply through PA just below the atomic asymptote. To ultimately create LiRb in the rovibronic ground state, we will likely need to find a lower lying PA resonance that will decay to the ground state. Alternatively, given the high PA rate to the $v = -3$ level of the $4(1)$ potential, it is possible that a two-step optical process could efficiently lead to ground state LiRb molecules as well.

We gratefully acknowledge support for the this project from the NSF (CCF-0829918), through an equipment grant from the ARO (W911NF-10-1-0243), and from university support through the Purdue OVPR AMO incentive grant. We would like to thank Dan Leaird for his assistance with these measurements.

\bibliography{masterlistfulljournal}

\end{document}